\documentclass{iau} 
\usepackage{graphicx}

\title[Gender equality activities in Astronomical Society of Japan]
{Gender equality activities\\
in Astronomical Society of Japan}

\author[Aya Bamba et al.]   
{Aya Bamba$^{1,2}$, Ayumi Asai$^3$, Ryohko Ishikawa$^4$,
Masayoshi Nobukawa$^5$, Hideko Nomura$^4$, Kosuke Sato$^6$,
Hisanori Furusawa$^4$, Mami Machida$^7$, 
\and Sachiko Okumura$^8$}

\affiliation{
$^1$Department of Physics, University of Tokyo, 7-3-1 Hongo, Bunkyo-ku, Tokyo 113-0033, Japan\\
email: {\tt bamba@phys.s.u-tokyo.ac.jp}\\[\affilskip]
$^2$Research Center for the Early Universe, School of Science, The University of Tokyo, 7-3-1 Hongo, Bunkyo-ku, Tokyo 113-0033, Japan\\[\affilskip]
$^3$Astronomical Observatory, Kyoto University, Sakyo, Kyoto, 606-8502, Japan\\[\affilskip]
$^4$National Astronomical Observatory of Japan, National Institutes of Natural Science, 2-21-1 Osawa, Mitaka, Tokyo 181-8588, Japan\\[\affilskip]
$^5$Department of Teacher Training and School Education, Nara University of Education, Takabatake-cho, Nara, 630-8528, Japan\\[\affilskip]
$^6$Graduate School of Science and Engineering, Saitama University, 255 Shimo-Ohkubo, Sakura, Saitama 338-8570, Japan\\[\affilskip]
$^7$Department of Physics, Faculty of Sciences, Kyushu University, 744 Motooka, Nishi-ku, Fukuoka, 819-0395, Japan\\[\affilskip]
$^8$Japan Women's University, 2-8-1 Mejirodai, Bunkyo-ku, Tokyo, 112-8681, Japan
}
\pubyear{2020}
\volume{358}  
\jname{Astronomy for equity, diversity and inclusion}
\editors{Lina Canas et al.}
\begin{document}

\maketitle

\begin{abstract}
The female ratio in science field, including astronomy,
is still quite low in Japan.
We, the Astronomical Society of Japan, keep making efforts
for the better gender balance.
In this paper, we summarize our statistics,
member’s thinking shown in our questionnaire,
the history and accomplishments of day-care system during annual meeting,
other activities, and so on.
\keywords{Keyword1, keyword2, keyword3, etc.}
\end{abstract}
\firstsection 

\section{Introduction}

The Astronomical Society of Japan (ASJ) was founded in 1908,
more than 100 years ago.
Now we have around 2000 full members and
the percentage of female researchers  is 13\%.
This is unfortunately below the avarage of other IAU member country cases.
We have studied the reason of this low female ratio.
The current state and member feeling of gender issue is summarized
in the following sections.

\section{From the statistics of ASJ full members}

\begin{figure}[thb]
\begin{center}
 \includegraphics[width=0.35\textwidth]{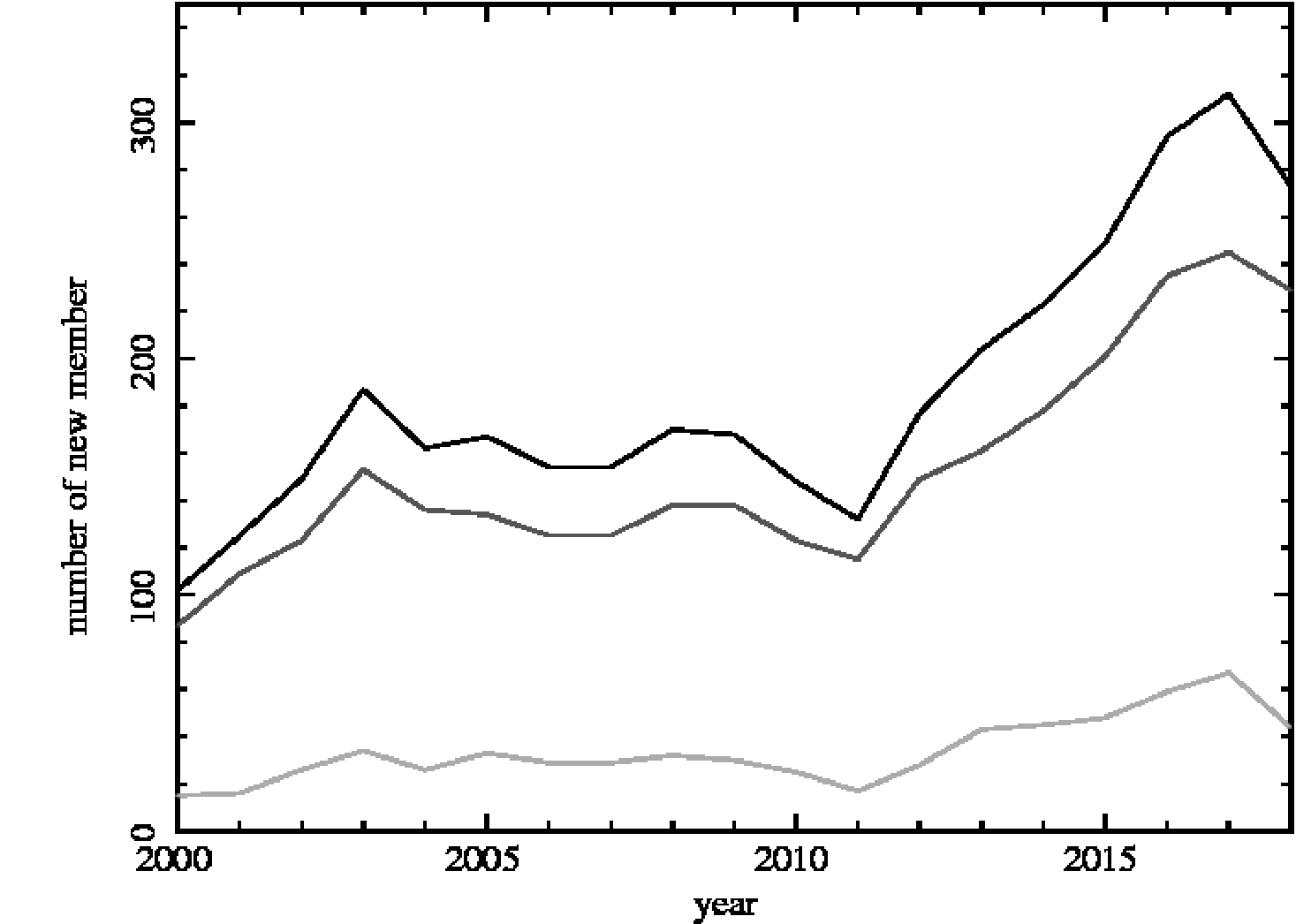} 
 \includegraphics[width=0.35\textwidth]{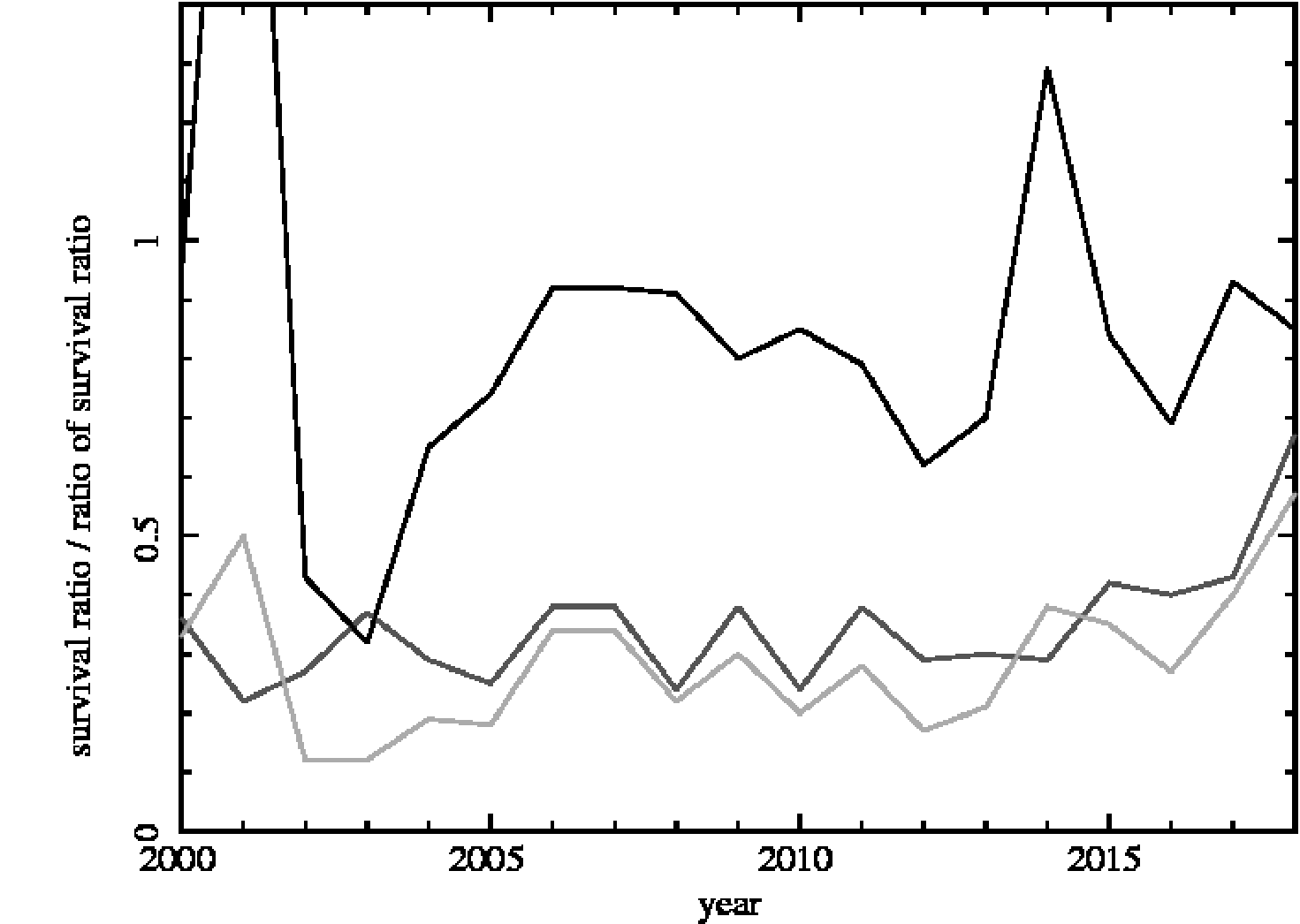} 
 \caption{Left: The number of new members in each year.
Black, dark gray, and light gray mean total, male, and female.
Right: the present number ratio who still in AsJ per joining year
(``survival ratio'').
Dark gray, light gray, and black mean male, female,
and female to male ratio.
}
   \label{fig1}
\end{center}
\end{figure}

ASJ have two kinds of personal membership,
``full membership'' for researchers including graduate students
and ``associate membership'' for amateurs.
Here we concentrate on statistics for members with full membership.
The left panel of figure~\ref{fig1} shows the number of ASJ new members
in each year (including both full and associate).
The number increases following evolution of astronomy.
The female ratio does not evolve so rapidly, keeping $\sim$18\%.
This number is larger than the total female ratio in ASJ,
implying there is ``leaky strow'' problem in the ASJ society.
The right panel of figure~\ref{fig1} shows the present ratio
who still in the ASJ society, sorted the joining year.
We define this number as ``survival ratio''.
Five years after they joined, the survival ratio is around 30\%,
suggesting it is tough to pursue academic careers for young researchers.
Moreover, the survival ratio is lower for female compared with male case
(see the right panel of Figure~\ref{fig1}).
The average female ratio is 82\% of male ratio.
It indicates that female researchers have special difficulties
to pursue academic careers compared with male researchers.
Similar results are also shown a survey in the physical society of Japan
\cite{nojiri2018}.

\section{From the survey in ASJ}

What makes special difficulties for female researchers?
Female researcher problem in ASJ is first revealed by
Dr. Mariko Kato, a female full member of ASJ;
she made first survey in ASJ in 1999.
The results are summarized in \cite{kato2000a} and \cite{kato2000b}. 

We have conducted another survey in ASJ.
to understand the present situation and compare the results with
\cite{kato2000a}.
The survey has done on April 2019,
twenty years after the survey by \cite{kato2000a}.
In order to study the difference of female and young generation situation,
we made questions as similar as those in \cite{kato2000a}.
The target is all ASJ members, and got $\sim$300 answers,
and the response rate is 14\%.
Note that the female response rate is 29\%,
implying that the problem is more recognized by female.
Here, we show several highlight results.

The first one is the experience getting strong objections 
when they joined master course.
Twenty years ago, 20\% female students got strong objections
to enter master course, whereas no male student got it.
On the other hand, the ratio of female students who got strong objections
becomes 10\% in 2019, around half of that in 20~years ago.
This result suggests that 
joining master course becomes easier for female students.
One interesting result is that 3\% of male now got strong objections,
which may be because of the recent difficulties to get tenured jobs.
For joining doctral couse, on the other hand,
we have no progress in these 20 years;
the female ratio who got disagree changed from 10\% to 13\%,
whereas male ratio changed from 6\% to 9\%.

\begin{figure}[htb]
\begin{center}
 \includegraphics[width=0.6\textwidth]{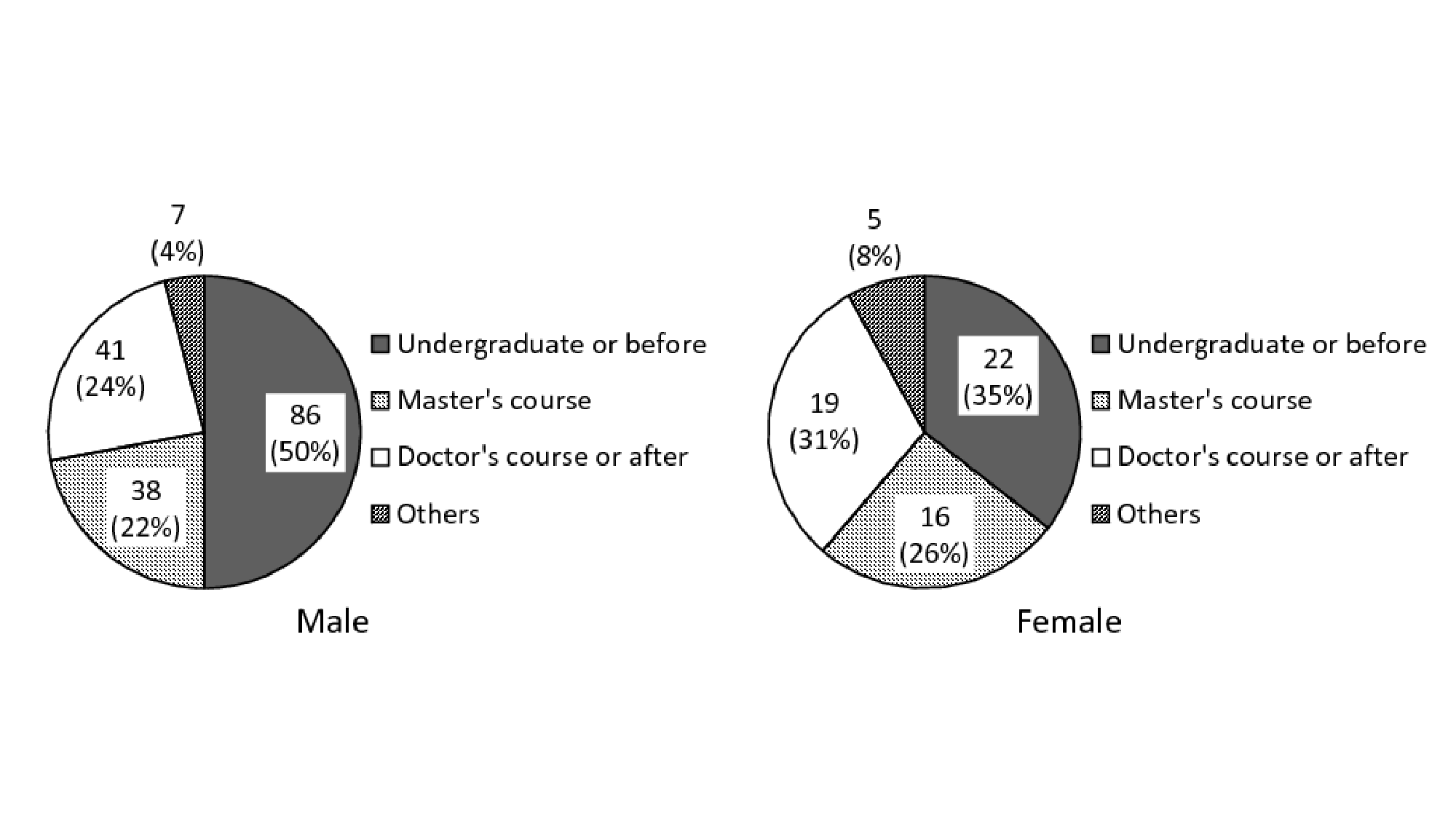} 
\vspace*{-4mm}
 \caption{The decision point distribution to be researchers
for male (left) and female (right).
}
   \label{fig2}
\end{center}
\end{figure}

Such an objection may affect the decision point to be professional researchers.
Figure~\ref{fig2} shows the age distribution
when they decide to be professional researchers.
One can see that the decision point is significantly later than male case.
It may imply that female can decide their career only after
they feel they can go together their career and life.
Such a delay can be one of the reasons for lower survival ratio.

\begin{figure}[hb]
\begin{center}
 \includegraphics[width=0.6\textwidth]{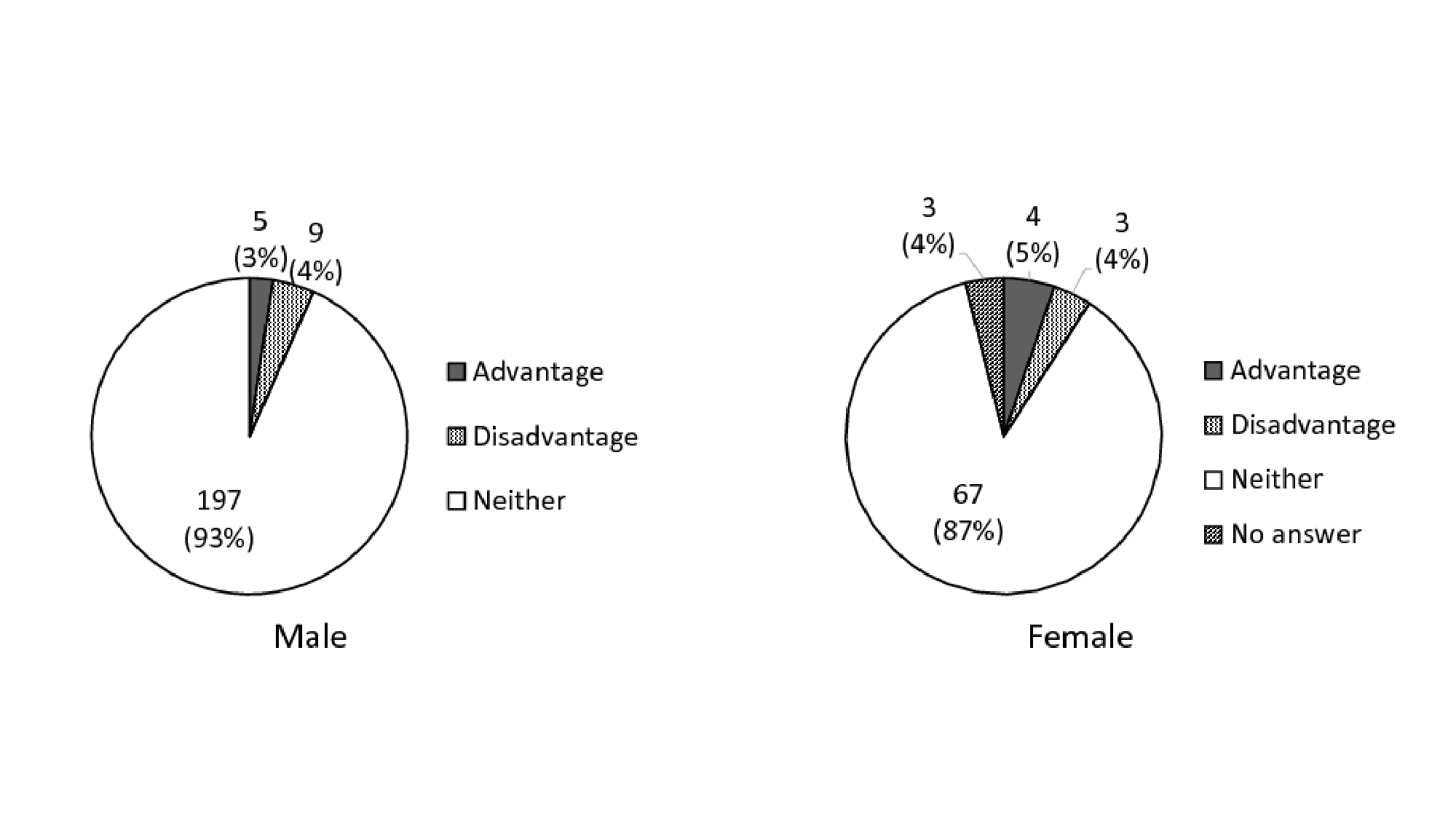} 
\vspace*{-4mm}
 \caption{The feeling whether your gender has advantage or disadvantage
to get tenured positions in astrophysics,
Left: male, Right: female.
}
   \label{fig3}
\end{center}
\end{figure}

Within such background,
some affirmative actions (i.e., job application only for female researchers)
have been introduced from $\sim$2010,
which is similar situation compared with other countries.
The number is still very limited in astrophysics;
we have checked the total number of tenured position calls
within these 10 years, and found that
only 4\% was devoted for the affirmative action positions.
This ratio is obviously insufficient to concave the difference of survival ratio
between female and male.
On the other hand, figure~\ref{fig3} shows the survey result
of the feeling whether their gender has advantage or disadavantage.
Surprisingly, both female and male feel female has advantage.
This can be just due to existence of affirmative action positions.
In order to accelerate gender equity in astronomical society in Japan,
more advertisement on the difficulty of female researchers
and why we need such actions.

\section{Actions for the future}

The ASJ have continued its effort for gender equity.
One of the most famous effort is setting temporal childcare
during annual meetings.
Dr. Kato introduced temporal childcare system in 1999 during an annual meeting.
It was the first case among Japanese academic societies.
In every annual meeting the childcare system helps several families.
Now this system spread over many societies from ASJ.
Interesting story is that
several young male researchers also used the first temporal childcare
with their children.
It means such an effort to help female (or minor) researchers
may be able to help young (or another kind of minor) researchers,
leading ASJ more attractive for younger generation.

\end{document}